\documentclass{article}
\usepackage[utf8]{inputenc}
\usepackage{graphicx}
\usepackage{amsmath,amssymb}
\usepackage[T1]{fontenc}
\usepackage{authblk}

\title{Topology optimization of freeform large-area metasurfaces}
\author[1]{Zin Lin\thanks{zinlin@mit.edu}}
\author[2]{Victor Liu}
\author[1,3]{Rapha\"{e}l Pestourie}
\author[1]{Steven~G.~Johnson}
\affil[1]{Department of Mathematics, Massachusetts Institute of Technology, Cambridge MA 02138, USA}
\affil[2]{San Francisco CA 94016, USA}
\affil[3]{John A. Paulson School of Engineering and Applied Sciences, Harvard University, Cambridge MA 02138, USA}

\date{February 2019}

\begin{document}

\maketitle

\begin{abstract}
We demonstrate optimization of optical metasurfaces over $10^5$--$10^6$ degrees of freedom in two and three dimensions, 100--1000+ wavelengths ($\lambda$) in diameter, with 100+ parameters per $\lambda^2$. In particular, we show how topology optimization, with one degree of freedom per high-resolution ``pixel,'' can be extended to large areas with the help of a locally periodic approximation that was previously only used for a few parameters per $\lambda^2$. In this way, we can computationally discover completely unexpected metasurface designs for challenging multi-frequency, multi-angle problems, including designs for fully coupled multi-layer structures with arbitrary per-layer patterns. Unlike typical metasurface designs based on subwavelength unit cells, our approach can discover both sub- and supra-wavelength patterns and can obtain both the near and far fields.
\end{abstract}


\section{Introduction}
We present a method for large-area ($\geq 100$ wavelengths $\lambda$) topology optimization (TO)~\cite{jensen2011topology,lalau2013adjoint,molesky2018inverse} of optical ``metasurfaces'' \cite{yu2011light, yu2014flat} with millions of degrees of freedom (DoF) determined automatically. Whereas previous metasurface design methods used only a few parameters per subwavelength unit cell~\cite{yu2014flat,aieta2012aberration,aieta2015multiwavelength,khorasaninejad2015achromatic,khorasaninejad2017visible,khorasaninejad2017achromatic,arbabi2017controlling,su2018advances,khorasaninejad2016metalenses}, TO allows us to consider thousands of parameters per unit cell, with much larger ($\sim 5\lambda$) unit cells in which both sub-wavelength and supra-wavelength features are discovered by making every ``pixel'' a DoF. Whereas previous TO methods in optics were restricted to computational domains $\lesssim 10\lambda$ amenable to brute-force simulation~\cite{sell2017large,lin2018topology}, we exploit a locally periodic approximation~\cite{aieta2012aberration,aieta2015multiwavelength,khorasaninejad2015achromatic,khorasaninejad2017visible,khorasaninejad2017achromatic,arbabi2017controlling,su2018advances,pestourie2018inverse} to combine many smaller simulations into a single optimization problem for a huge surface (Sec.~\ref{sec:formulation}). We validate example designs for high numerical-aperture (NA) multi-frequency/multi-angle lenses in 2d against full-wave simulations of the entire domain (Sec.~\ref{sec:conc} and \ref{sec:chrome}). We also present example 3d designs for monochromatic high-NA lenses with $\sim 10^6$ DoF (Sec.~\ref{sec:3d}), enabled by an efficient massively parallel implementation combining thousands of RCWA (rigorous coupled-wave analysis) unit-cell solutions~\cite{moharam1981rigorous, Liu20122233} and fast adjoint-method~\cite{jensen2011topology, molesky2018inverse} gradient computations. Because our method employs the full scattered field for each surface unit, as opposed to a single phase or amplitude in previous works~\cite{yu2014flat,aieta2012aberration,aieta2015multiwavelength,khorasaninejad2015achromatic,khorasaninejad2017visible,khorasaninejad2017achromatic,arbabi2017controlling,su2018advances}, we could also apply our method to near-field or guided-wave optimization~\cite{Perez-Arancibia:18} (Sec.~\ref{sec:final}).

Metasurfaces offer ultra-compact flat-optics alternatives for traditional bulky systems used in lensing, beam-shaping, holography and beyond~\cite{capasso2018future}. The power and versatility of a metasurface resides in aperiodically patterned nanophotonic features covering hundreds or thousands of wavelengths in diameter. The term ``meta" often refers to extremely subwavelength features that may be modeled by effective surface impedances~\cite{pfeiffer2013metamaterial}, but in this work we consider both sub- and supra-wavelength features and do not employ any effective-medium approximation. Understandably, a meta-device poses a more challenging design problem than a traditional optical system since it requires an understanding and control of electromagnetic interactions within a vast number of nanostructures. Typically, such interactions are handled by a locally periodic approximation (LPA) in which the metasurface is divided into computationally tractable independent unit cells with periodic boundary conditions~\cite{yu2014flat,aieta2012aberration,aieta2015multiwavelength,khorasaninejad2015achromatic,khorasaninejad2017visible,khorasaninejad2017achromatic,arbabi2017controlling,su2018advances}. One extreme limit of LPA is scalar diffraction theory~\cite{born2013principles}, in which the surface is treated as locally \emph{uniform}; this is often used to design diffractive surfaces~\cite{kim2012design} but is unlikely to accurately model subwavelength patterns. Meanwhile, the geometric library which provides the ``meta-elements" is made up of primitive shapes such as subwavelength cylindrical or rectangular pillars which can be rapidly modeled (under LPA) by an electromagnetic solver such as RCWA~\cite{yu2014flat,aieta2012aberration,aieta2015multiwavelength,khorasaninejad2015achromatic,khorasaninejad2017visible,khorasaninejad2017achromatic,arbabi2017controlling,su2018advances}. However, it may become increasingly difficult and ultimately infeasible to leverage only simple primitive geometries for more complex problems such as broadband achromatic focusing~\cite{chen2018broadband,wang2018broadband,shrestha2018broadband} or controlled angular dispersion~\cite{kamali2017angle} which impose stringent demands on the local phase shifts provided by the meta-elements.         

Topology optimization (TO) is a large-scale computational technique that can handle an extensive design space, considering the dielectric permittivity at every spatial point as a DoF~\cite{jensen2011topology,lalau2013adjoint,molesky2018inverse}. In contrast to the heuristic search routines regularly employed by the photonic community such as genetic algorithms~\cite{mitchell1998introduction} or particle swarm methods~\cite{kennedy2011particle}, TO employs gradient-based optimization techniques to explore hundreds to billions of \emph{continuous} DoFs. Such a capability is made possible by a rapid computation of gradients (with respect to all the DoFs) via adjoint methods~\cite{strang2007computational}, which, in the context of electromagnetic inverse design, involve just one additional solution of Maxwell's equations~\cite{molesky2018inverse}. In fact, such techniques have been gaining traction lately in the field of photonic integrated circuits, especially for the design of compact modal multiplexers and converters~\cite{piggott2015inverse}. Only recently, TO-based inverse design methods have been extended to metasurfaces, particularly in the context of freeform meta-gratings~\cite{sell2017large} and metalenses with angular phase control~\cite{lin2018topology}. While such applications reveal TO as a very promising tool for realizing increasingly sophisticated meta-devices, they have been limited to small computational domains $\lesssim 10\lambda$. In this paper, we present a combination of TO and LPA to efficiently design large-area metasurfaces with enhanced functionalities. 

\section{Formulation}
\label{sec:formulation}
\begin{figure}[b]
    \centering
    \includegraphics[scale=0.4]{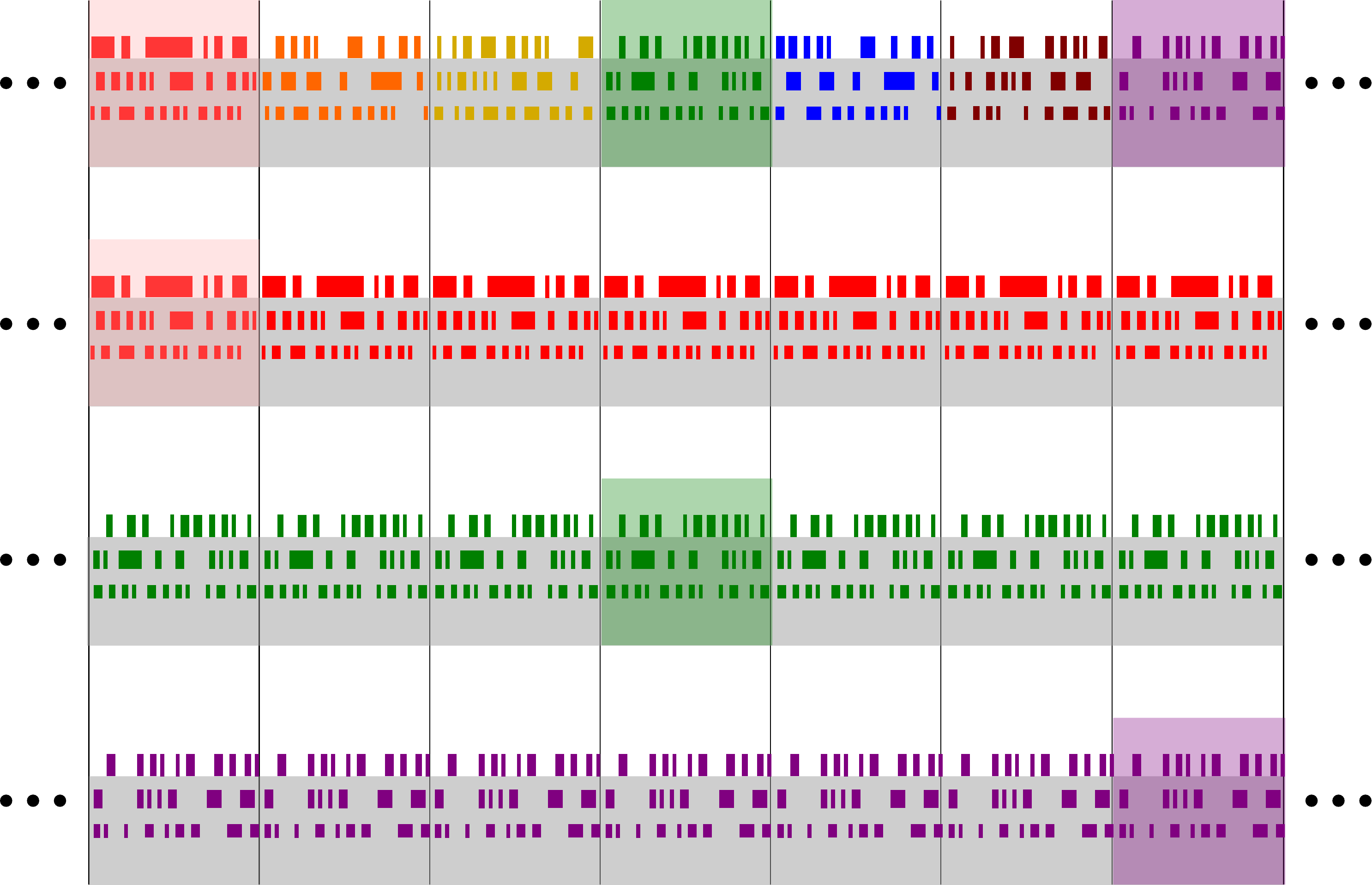}
    \caption{An arbitrary aperiodic multi-layered meta-structure (top) is approximated by solving a set of periodic scattering problems (bottom), one for each unit cell (shaded areas), to obtain the approximate near fields over the entire metasurface.}
    \label{fig0}
\end{figure}
Because we employ a general optimization framework, we have great flexibility in what function of the electromagnetic fields to optimize (our ``objective'' function $f$)~\cite{pestourie2018inverse}. For many light-focusing problems, it is convenient to simply maximize the electric-field intensity at a focal point $\mathbf{r_0}$~\cite{pestourie2018inverse}:
\begin{align}
    \max_{\big\{\bar{\epsilon}(\mathbf{r})\big\}} f\left( \mathbf{E}; \bar{\epsilon} \right) &= \Bigg| \int \mathbf{G_0}\left(\mathbf{r_0},\mathbf{r_s}\right) \cdot \mathbf{E}(\mathbf{r_s})~d\mathbf{r_s} \Bigg|^2 \label{eq1} \\
    0 &\le \bar{\epsilon} \le 1. \notag
\end{align}
Here, the set of DoFs $\big\{\bar{\epsilon}(\mathbf{r})\big\}$ is related to the position-dependent dielectric profile via $\epsilon(\mathbf{r}) = \left( \epsilon_\text{st} -
\epsilon_\text{bg} \right) \bar{\epsilon}\left(\mathbf{r}\right) + \epsilon_\text{bg}$, where $\epsilon_\text{st (bg)}$ denotes the relative permittivity of the structural (background) dielectric material. Whereas $\bar{\epsilon}$ is a \emph{continuous} DoF allowed to have intermediate values between 0 and 1, we employ Heaviside projection filters~\cite{jensen2011topology} to ensure that the final optimal design is binary, i.e., $\bar{\epsilon}_\text{optimal} \in \{0,1\}$. In this work, we consider only the binary filter to theoretically illustrate our design method. More generally, a variety of additional geometric filters and constraints can be straightforwardly incorporated into the formulation, including regularization filters and curvature constraints~\cite{jensen2011topology,zhou2015minimum,piggott2017fabrication} to impose design robustness and minimum feature sizes conducive to fabrication. The objective function $f$ represents the far-field intensity obtained by convolving the free-space Green's function $\mathbf{G_0}$ with the near-field $\mathbf{E}$ at some reference plane $\mathbf{r_s}$ over the entire metasurface. (To be precise, the free-space Green's function is convolved with equivalent surface currents which are given by the tangential field components~\cite{emsrc}.) Here, $\mathbf{E}$ is the steady-state solution to the frequency-domain Maxwell's equation:
\begin{align}
    \nabla \times \mu^{-1} \nabla \times \mathbf{E} - \omega^2 \epsilon \mathbf{E} = -i \omega \mathbf{J} 
\end{align}
in response to the incident current $\mathbf{J}$ at a frequency $\omega$ where we set the magnetic permeability $\mu=1$ for typical optical materials. To efficiently model a large device area, we use the locally periodic approximation~\cite{pestourie2018inverse}, in which the metasurface is broken up into multiple smaller cells, each of which is independently simulated with periodic boundary conditions~(Fig.~\ref{fig0}). The total near-field $\mathbf{E}$ is, therefore, approximated by the set of disjoint intra-cell electric fields which must be engineered such that they all constructively add up at the focal point. The optimization typically involves thousands of DoFs which must be efficiently handled by a numerical algorithm. Within the scope of TO, this requires efficient calculations of the derivatives ${\partial f \over \partial \bar{\epsilon} }$ at every pixel point in the device region, which we obtain by an adjoint method~\cite{molesky2018inverse,strang2007computational}:
\begin{align}
    \nabla_{\bar{\epsilon}} f(\mathbf{r}) &= 2 \omega^2 (\epsilon_\text{st}-\epsilon_\text{bg}) ~\operatorname{Re}\Bigg\{ g^*~\mathbf{\Tilde{E}}(\mathbf{r}) \cdot \mathbf{E}(\mathbf{r}) \Bigg\} \\
    g &= \int \mathbf{G_0}\left(\mathbf{r_0},\mathbf{r'}\right) \cdot \mathbf{E}(\mathbf{r'})~d\mathbf{r'},
\end{align}
where the adjoint field $\mathbf{\Tilde{E}}$ is evaluated by solving an additional Maxwell's equation:
\begin{equation}
        \nabla \times \mu^{-1} \nabla \times \mathbf{\Tilde{E}} - \omega^2 \epsilon \mathbf{\Tilde{E}} = \mathbf{G_0}\left(\mathbf{r_0},\mathbf{r}\right)~\mathbf{\delta}(\mathbf{r}-\mathbf{r_s}).
\end{equation}
We emphasize that our framework is entirely different from the common approach where an ideal phase profile to be realized is approximately fitted with zeroth-order phase shifts allowed by primitive scattering elements, each residing within a sub-wavelength unit cell~\cite{yu2014flat,aieta2012aberration,aieta2015multiwavelength,khorasaninejad2015achromatic,khorasaninejad2017visible,khorasaninejad2017achromatic,arbabi2017controlling,su2018advances}. In contrast, our approach greatly broadens the structural design space by considering non-intuitive shapes and forms as well as supra-wavelength unit cells within which the amplitude and phase of the electric field may vary considerably and must be fully taken into account during optimization.

\subsection{Numerical Implementation}
To ensure a speedy computation, we numerically implement the optimization problem~(\ref{eq1}) in C. Each of the cells is either solved by a C implementation of the finite-difference frequency-domain method (FDFD)~\cite{shin2012choice} or the rigorous coupled-wave analysis (RCWA)~\cite{Liu20122233}. To model a large surface area rapidly, all the cells are independently simulated in an ``embarrassingly parallel'' fashion using the message-passing interface (MPI) library~\cite{gropp1999using,pacheco2011introduction}. The simulated results are then consolidated to compute the global objective and gradients. The structural update during optimization is provided by a standard gradient-based nonlinear-optimization algorithm~\cite{svanberg1987method,svanberg2002class} with a free-software implementation~\cite{johnson2014nlopt}. We note that the discretization error in the gradients arising from the discrepancy between the formal adjoint method and the numerical Maxwell's solver (such as RCWA) is practically negligible ($\lesssim 0.01\%$).

\section{Applications}
The major advantage of a large-scale computational approach consists in its flexibility for handling multi-objective problems, such as those prescribing the behavior of a single metasurface for a set of frequencies and incident current conditions $\{\omega_i,\mathbf{J}_i\}$. In such cases, the optimization problem can be easily extended by the maximin formulation:
\begin{align}
    \max_{\bar{\epsilon}} \min_i \Big\{ f_{\omega_i,\mathbf{J}_i} \Big\},
\end{align}
which can be recast into an equivalent differentiable form~\cite{boyd2004convex}:
\begin{align}
    \max_{\bar{\epsilon},t}~&t \\
    t - f_{\omega_i,\mathbf{J}_i} &\leq 0.
\end{align}

\subsection{Metalens Concentrator}
\label{sec:conc}

\begin{figure}
    \centering
    \includegraphics[scale=0.35]{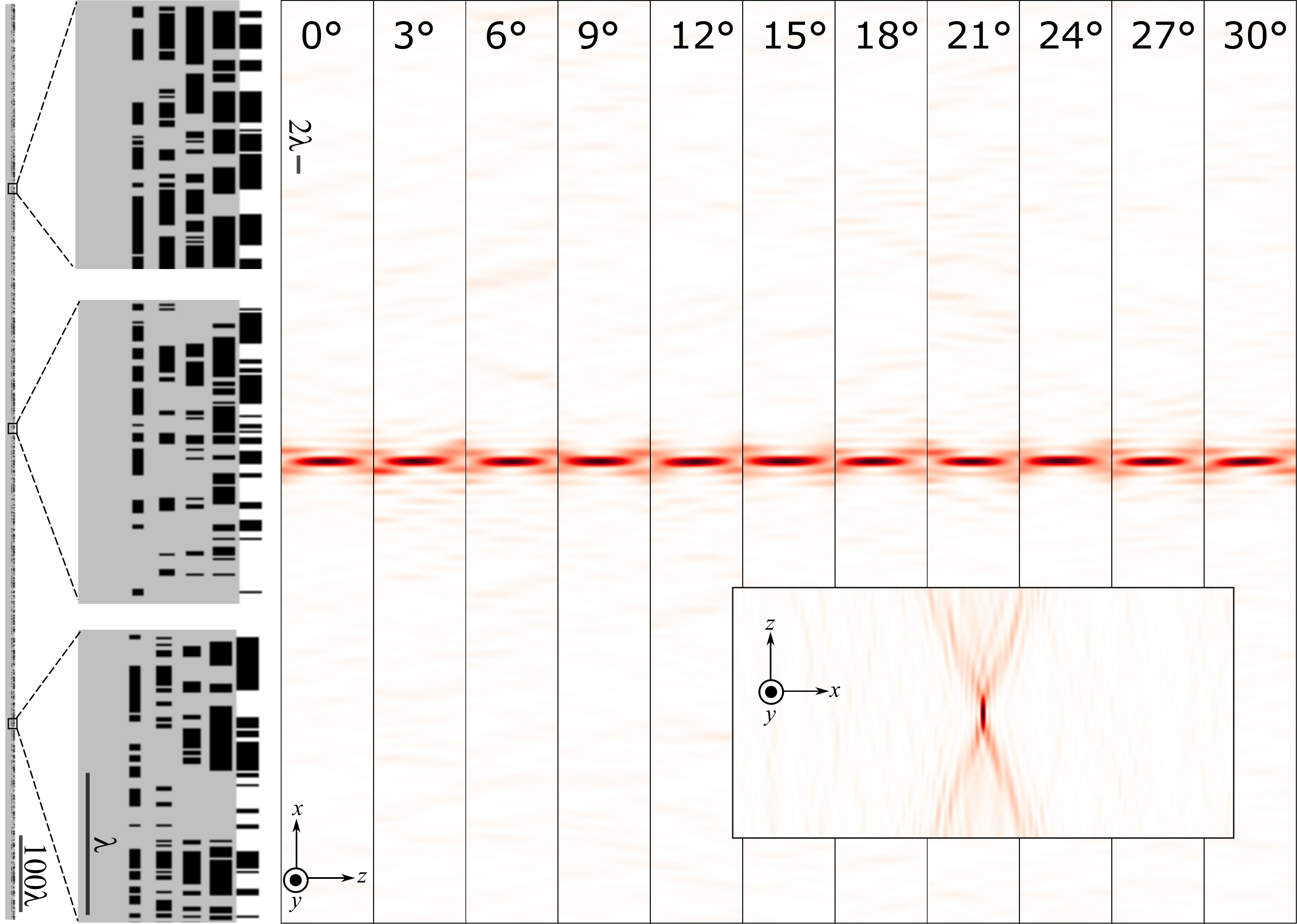}
    \caption{Multi-layered 2D metalens concentrator (NA=0.51) which can combine 11 incident angles to a single focus. The lens consists of five layers of TiO$_2$, with thicknesses of 0.10$\lambda$, 0.14$\lambda$, 0.16$\lambda$, 0.20$\lambda$ and 0.24$\lambda$ respectively, situated above and within the silica substrate. Three different portions of the lens have been magnified for an easy viewing of the device geometry; note the scale bars. Full-wave simulations of the entire structure reveal  diffraction-limited focusing for the 11 incident angles with average transmission efficiency of $\approx 40$\%. }
    \label{fig1}
\end{figure}
To demonstrate the efficacy of our design technique, we optimize a multi-layered 2D metalens (Fig.~\ref{fig1}) that can concentrate incoming light at 11 incident angles $\{\theta~(^\circ)=3i ~\big|~ i=0,1,...,10\}$ to the same focus. The lens consists of five layers of TiO$_2$ (refractive index $n \approx 2.4$), four of which are buried in silica ($n \approx 1.5$). The diameter of the lens is $D = 1200\lambda$, where $\lambda$ is the operational wavelength, and the focal length is $F = 1000\lambda$, corresponding to a rather high NA of 0.51. For simplicity, we consider the electric field to be polarized along the $y$ axis. During the optimization, the entire lens is broken up into 240 cells, each of which is $5\lambda$ long, for $3 \times 10^5$ DoFs in total, while the entire computation is parallelized over 1200 CPUs. While we perform the optimization with the locally periodic approximation, we validate the optimized result with rigorous full-wave FDTD simulations~\cite{oskooi2010meep}~(Fig.~\ref{fig1}) in which we compute the actual electric fields over the entire metasurface without any uncontrolled approximation. The lens is shown to exhibit diffraction-limited focusing at all the optimized angles with the full-width-at-maximum (FWHM) of $\sim \lambda$ while the average transmission efficiency is found to be $T \approx 40\%$ . The average efficiency of $40\%$ seems to violate Lorentz reciprocity and, in particular, the concentration bound $\leq {1 \over 11}$ predicted by~\cite{zhang2018scattering}. However, there is no such violation because, although the output field profiles have the same prominent peak at the optical axis, they are never entirely alike for any two input angles due to the weak but distinct diffracting patterns away from the axis.

\begin{figure}
    \centering
    \includegraphics[scale=0.5]{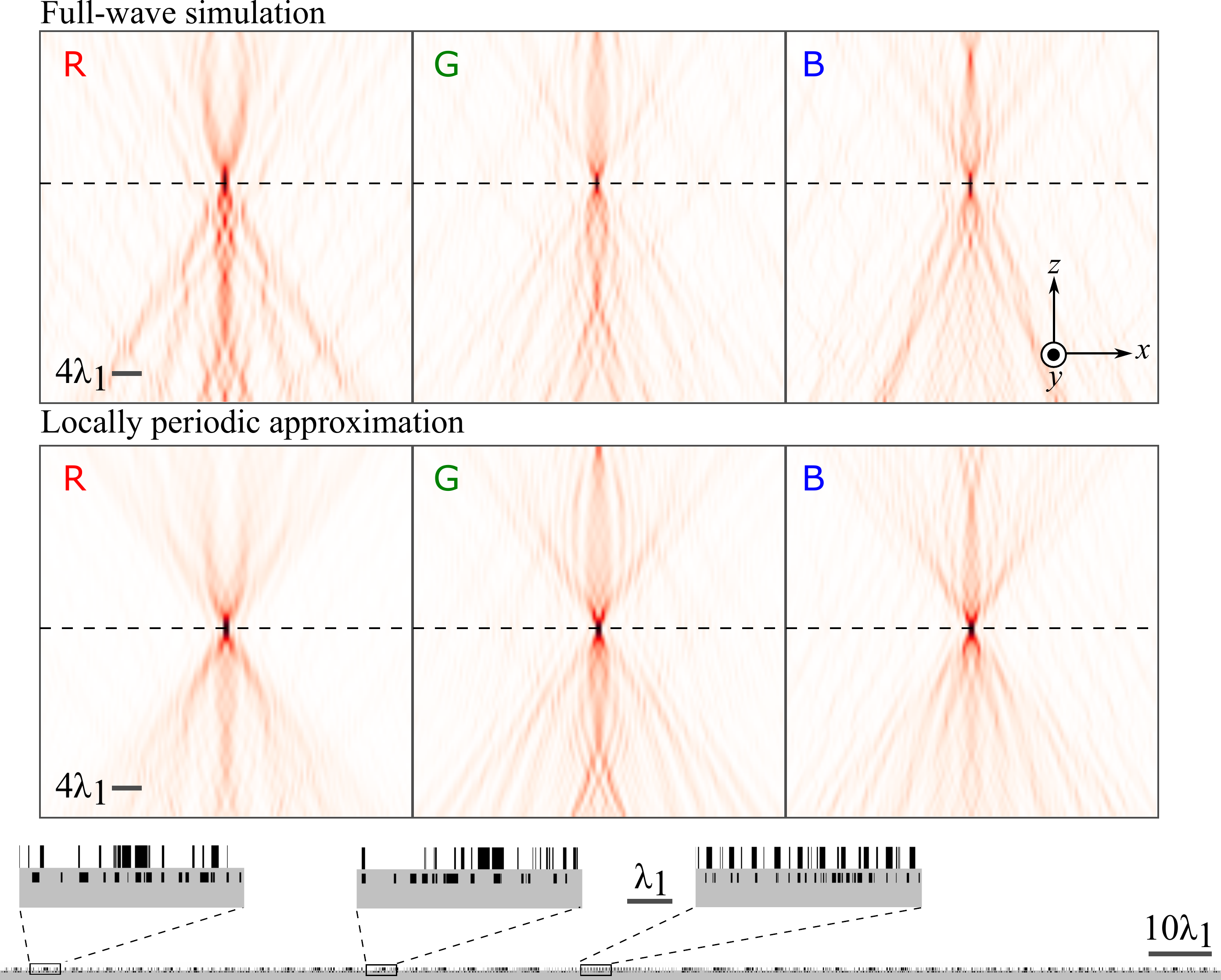}
    \caption{Multi-layered 2D metalens (NA=0.45) chromatically corrected at the wavelengths $\lambda_1=650~\mathrm{nm}$, $\lambda_2=540~\mathrm{nm}$ and $\lambda_3=470~\mathrm{nm}$. The lens consists of two layers of TiO$_2$, with thicknesses of 0.5$\lambda_1$ and 0.2$\lambda_1$ respectively, situated above and within the silica substrate. Three different portions of the lens have been magnified for an easy viewing of the device geometry; note the scale bars. The far field profiles are obtained by full-wave simulations of the entire structure (the upper panel) and by locally periodic approximation (the lower panel), exhibiting at- or near-diffraction limited focusing.}
    \label{fig2}
\end{figure}

\subsection{Multi-wavelength Focusing}
\label{sec:chrome}
Another example we consider is a partially achromatic 2D metalens (Fig.~\ref{fig2}) with discrete chromatic aberration corrections at three wavelengths $\{\lambda_1,\lambda_2,\lambda_3\}$ with $\lambda_2 = \lambda_1/1.2$ and $\lambda_3 = \lambda_1/1.38$. In particular, one may choose $\lambda_1=650~\mathrm{nm}$, $\lambda_2=540~\mathrm{nm}$ and $\lambda_3=470~\mathrm{nm}$, corresponding to red, green and blue (RGB) wavelengths. The lens consists of two layers of TiO$_2$ ($n \approx 2.4$), one of which is buried in silica ($n \approx 1.5$). The diameter of the lens is $D = 200\lambda_1$ and the focal length is $F = 100\lambda_1$, corresponding to $\mathrm{NA} = 0.71$. During the optimization, the entire lens is broken up into 40 cells, each of which is $5\lambda$ long, for $1.6 \times 10^4$ DoFs in total, while the entire computation is parallelized over just 20 CPUs. At the desired frequencies, the lens is shown to focus a normally incident $y$-polarized beam at or near the diffraction limit with the FWHMs of $\approx 0.71 \lambda_1,~0.72 \lambda_2$ and $0.76 \lambda_3$ respectively and with an average transmission efficiency of $\sim 71\%$. Recently, broadband achromatic metalenses have been gaining widespread attention for full color imaging; although the state-of-the-art designs continuously cover the full visible spectrum, they are mostly limited to smaller lens diameters and/or low NA~\cite{chen2018broadband,wang2018broadband,shrestha2018broadband}. Our results suggest that multi-layer designs enabled by TO could alleviate many of these limitations by allowing more complex geometries to be explored.

Generally, a high-NA multi-wavelength design is a challenge for LPA because the design rapidly varies across the surface, resulting in the scattering ``noise'' visible in a full-wave simulation of the entire structure but not captured by the conjoined LPA solution (Fig.~\ref{fig2}), in contrast to the multi-angle case in Fig.~\ref{fig1}. One consequence is that the focusing efficiencies at the three wavelengths (defined as the integrated field intensity around the focal spot divided by the total intensity over the image plane~\cite{khorasaninejad2016polarization}) are found to be $\sim51\%,~50\%$  and $45\%$ while LPA predicts $\sim78\%,~55\%$ and $60\%$ respectively. Note that our objective function~(\ref{eq1}) merely maximizes the intensity at the focal spot but does not necessarily suppress stray diffractions elsewhere which become apparent even in LPA predictions (Fig.~\ref{fig2}). If desired, the optimal design can be further refined by improving the LPA wavefront by minimizing a phase error~\cite{lin2018topology} or wavefront error objective~\cite{pestourie2018inverse}. At the same time, deviations from LPA may be mitigated by considering larger metasurfaces where LPA is more appropriate, by including higher-order corrections to LPA~\cite{Perez-Arancibia:18}, or by incorporating optimization constraints designed to enforce the validity of LPA. For example, the optimization process may be augmented with constraints restricting the structural variations between neighboring cells or specifying a bound on the physical residual:
\begin{align}
    r(\epsilon)=\Big|\nabla \times \mu^{-1} \nabla \times \mathbf{E_\text{LPA}} - \omega^2 \epsilon \mathbf{E_\text{LPA}} + i \omega \mathbf{J} \Big|^2
\end{align}
where the double curl operator is applied over the whole surface without LPA. Note that evaluating the residual $r$ requires just a single matrix-vector multiplication, albeit involving a very large and sparse matrix, and can be made relatively cheap and fast when parallelized over several cores.

\subsection{Three-dimensional Examples}
\label{sec:3d}

\begin{figure}
    \centering
    \includegraphics[scale=0.17]{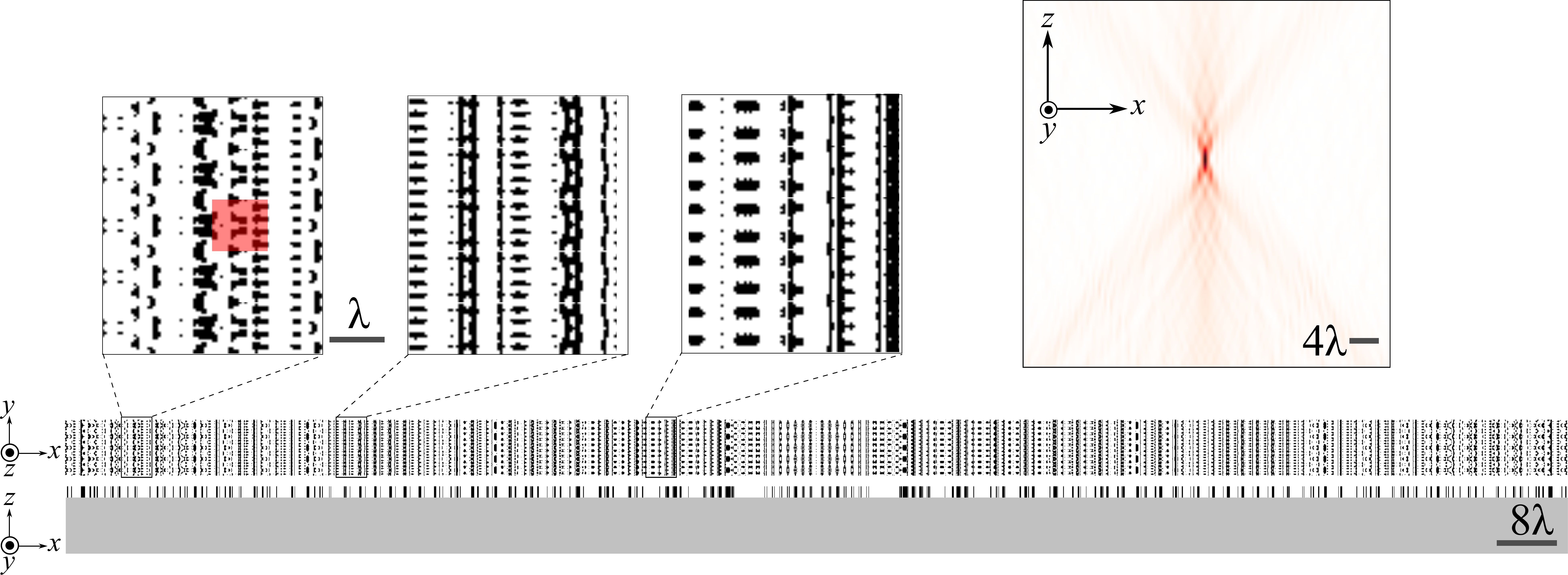}
    \caption{Monochromatic 3D cylindrical metalens (NA=0.71) with aperiodic cells along the $x$ axis. The lens consists of a single TiO$_2$ layer above the silica substrate. Three different portions have been magnified for easy viewing; note the scale bars. The shaded area shows an example of a $\lambda \times \lambda$ cell. A full-wave simulation of the entire structure shows the diffraction-limited focusing.}
    \label{fig3}
\end{figure}

Next, we turn to proof-of-concept 3D examples. First, we consider a collection of 3D cells, each of which has an area $L_x \times L_y = \lambda \times \lambda$. The cells are repeated along $y$ but are allowed to differ from each other along $x$. Fig.~\ref{fig3} shows the corresponding ``cylindrical'' lens (NA=0.71) comprising such cells which focuses a normally-incident $y$-polarized beam to a focal line running parallel to the $y$-axis. The lens, though periodic along $y$, shows complex geometric variations along both $x$ and $y$ within each cell. A rigorous FDTD simulation of the whole lens shows the focusing behavior with a transmission efficiency of 75\%. Although the design of a monochromatic lens, no matter the geometric complexity, is only useful to illustrate the method, we note that our framework enables a relatively cheap and fairly quick computation using as few as 25 CPUs over a span of a few minutes while considering the entire structure with as many as 200 cells and $4 \times 10^4$ DoFs in total.

\begin{figure}
    \centering
    \includegraphics[scale=0.5]{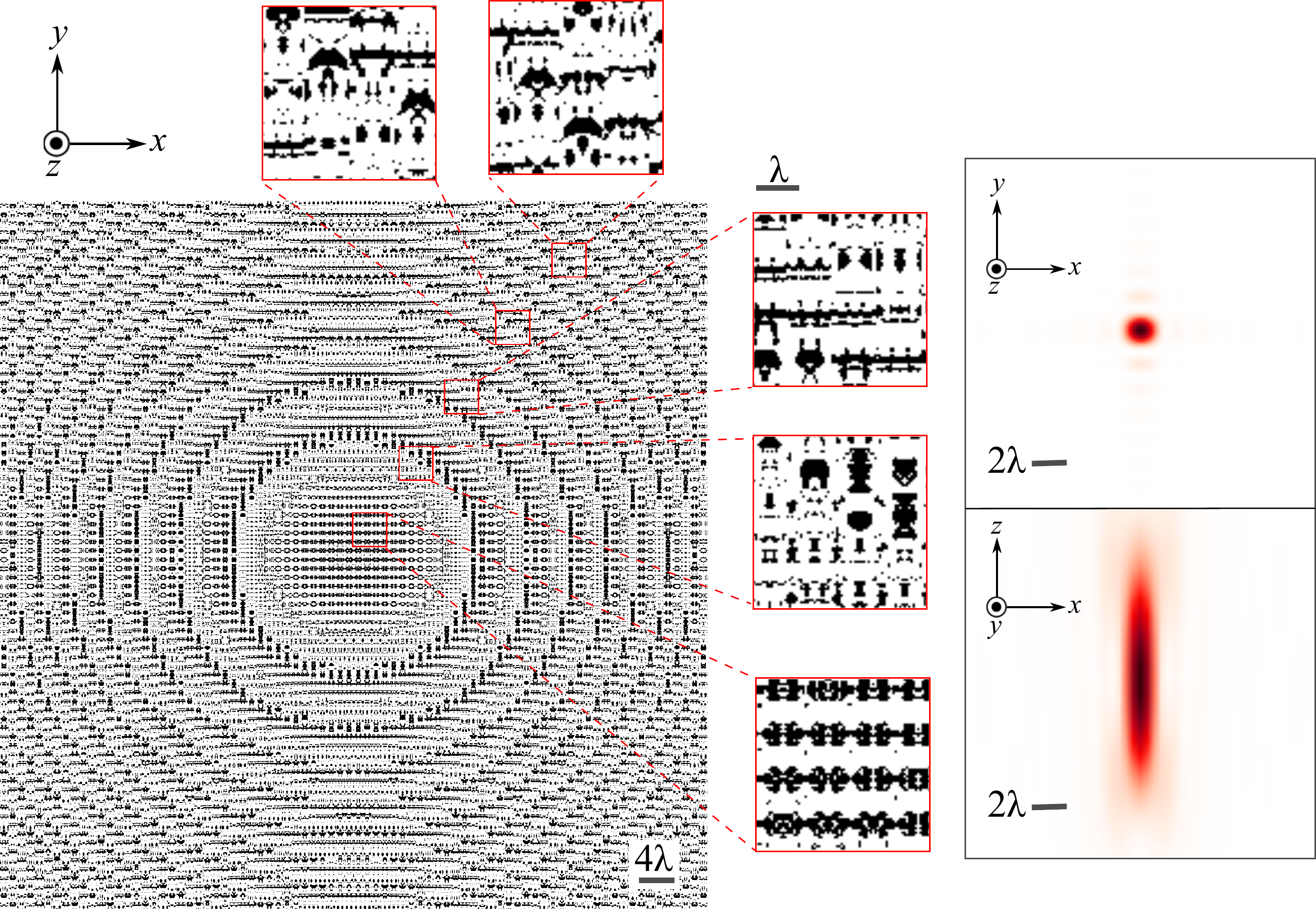}
    \caption{Monochromatic 3D metalens (NA=0.37). A few portions of the lens have been magnified for easy viewing; note the scale bars. The lens consists of a single TiO$_2$ layer above the silica substrate. The far field profile is obtained by locally periodic approximation, showing diffraction-limited focusing.}
    \label{fig4}
\end{figure}

Lastly, we consider the full 3D lens (Fig.~\ref{fig4}) with cells varying along both $x$ and $y$. The lens has 6400 $\lambda \times \lambda$ cells, corresponding to NA=0.37, with $6.4 \times 10^5$ DoFs in total. Again, our framework enables a cheap and speedy design of the entire lens, utilizing 40 CPUs and lasting only a few minutes. A major computational hurdle actually arises \emph{after} the optimization: given the enormous size, the lens cannot be readily simulated via FDTD using a few CPUs. Instead, we provide an approximate far-field profile computed from the LPA combination of the unit-cell simulations, which has been shown to give excellent accuracy for monochromatic designs~\cite{khorasaninejad2016metalenses} and good accuracy even for the complex multi-frequency designs in Sec.~\ref{sec:chrome}.

These preliminary results suggest promising ways to scale up the 3D metasurface design. In particular, the computational workload can be reduced by orders of magnitude via rotational symmetry, specifically, by considering just a single row of cells and rotating it to fully cover the entire surface. Such an approach is, in fact, a more common practice in metalens design~\cite{yu2014flat,aieta2012aberration,aieta2015multiwavelength,khorasaninejad2015achromatic,khorasaninejad2017visible,khorasaninejad2017achromatic,arbabi2017controlling,su2018advances,khorasaninejad2016metalenses} rather than considering all the cells as we did here. Also, polarization insensitivity could be ensured by further imposing appropriate symmetries, such as $C_{4v}$ within each cell. Combined with such symmetry-related reductions, an access to several hundreds or low thousands of CPUs (readily available on institutional supercomputers or commercial cloud computing services) amply allow for the design of high-NA wide-area metalenses with greatly enhanced functionalities including broadband achromaticity and large-angle aberration corrections.

\section{Conclusion and Outlook}
\label{sec:final}
We have provided an efficient computational framework for the inverse design of large-area freeform metasurfaces. We have also presented various examples including multi-angle and multi-wavelength metalens designs that demonstrate the versatility and power of our method. We expect that a large-scale optimization technique like ours may become indispensable for tackling challenging problems which inherently call for a large design space with multiple layers. In particular, our method may greatly benefit the design of a polarization-insensitive large-area high-NA \emph{single-piece} metalens with chromatic and achromatic aberration corrections over the entire visible spectrum and over a large field of view. On the other hand, our framework need not be limited to metalens design. In fact, the particular objective we presented above might be more naturally suited for three-dimensional sculpting of the far-field intensity (or thin-film holography). Another intriguing application particularly suitable for the RCWA-based optimization framework is to inverse-design the far field between two meta-devices separated by thick (many-wavelengths) spacing layers. Ultimately, one would like to address the most challenging problem of electromagnetic design: to intimately manipulate the near-field physics of large aperiodic structures in which ultra-rich modal interactions involving both extended and localized resonances may offer a superb playground for exploring novel phenomena and devising next-generation technologies such as solid-state Lidar systems~\cite{poulton2017coherent}, strongly enhanced light-matter interactions and nonlinear processes~\cite{pick2017general,lin2016enhanced,rivera2016shrinking,lin2016cavity,venkataram2017unifying}, near-field radiative heat transfer~\cite{rodriguez2011frequency}, and opto-nanomechanical engineering~\cite{reid2017photon,lee2017computational}. While LPA offers asymptotic advantages in the limit of ``adiabatic'' nearly periodic surfaces, our optimization framework also provides a promising pathway for handling a wider variety of large aperiodic systems by incorporating additional constraints (Sec.~\ref{sec:chrome}) or by augmenting LPA with domain decomposition concepts~\cite{chan1994domain} including using the conjoined LPA solution as a preconditioner, using other boundary conditions (e.g. mirror boundaries), and using overlapping domains~\cite{zhao2018accelerating,tao2016integral}.

\end{document}